\documentclass[11pt,letterpaper]{article}
\usepackage[top=0.85in,left=1in,right=1in,footskip=0.75in,marginparwidth=1in]{geometry}

\usepackage[T1]{fontenc}

\usepackage{graphicx}

\usepackage [english]{babel}
\usepackage [autostyle, english = american]{csquotes}
\MakeOuterQuote{"}

\usepackage{nameref}
\usepackage{hyperref}

\usepackage[dvipsnames]{xcolor}
\usepackage{cleveref}
\newcommand\myshade{85}
\colorlet{mylinkcolor}{violet}
\colorlet{mycitecolor}{YellowOrange}
\colorlet{myurlcolor}{Aquamarine}

\hypersetup{
  linkcolor  = mylinkcolor!\myshade!black,
  citecolor  = mycitecolor!\myshade!black,
  urlcolor   = myurlcolor!\myshade!black,
  colorlinks = true,
}

\usepackage{parskip}%
\usepackage{caption}
\usepackage{subcaption}

\usepackage[
backend=biber,
style=authoryear-icomp,
sorting=none, %
maxcitenames=1,
maxbibnames=9,
minbibnames=5, %
]{biblatex}
\usepackage{xcolor}

\title{Tracking e-cigarette warning label compliance \\on Instagram with deep learning}
\author{Chris J. Kennedy, Julia Vassey, Ho-Chun Herbert Chang,\\
Jennifer B. Unger, Emilio Ferrara}%

\begin{document}

\maketitle

\section{Introduction}

The proportion of the U.S. high school students who report using e-cigarettes (aka vaping devices) declined in 2020 during the COVID-19 pandemic. In 2020, 19.6\% of high school students (3.02 million) reported current e-cigarette use, compared to 27.5\% (4.11 million) of high students who reported using e-cigarettes in 2019  \parencite{wang2020cigarette}. However, despite this recent decline, during 2019 - 2020, the use of youth-appealing, low-priced disposable e-cigarette devices increased approximately 1,000\% (from 2.4\% to 26.5\%) among high school current e-cigarette users \parencite{wang2020cigarette}. In addition, more than eight in 10 teenage e-cigarette users reported consuming flavored e-cigarettes \parencite{wang2020cigarette}. E-cigarettes can harm the adolescent brain and increase susceptibility to tobacco addiction \parencite{us2016cigarette, fraga2019dangers, wang2019tobacco}. As a result, youth who use e-cigarettes are more likely to subsequently try combustible cigarettes \parencite{us2018fda}. Beyond addiction, e-cigarettes pose a risk of breathing difficulties, inflammatory reactions, lowered defense against pathogens and lung diseases \parencite{redfield2019cdc, fda2019dangers}. 
Exposure to visual posts featuring e-cigarette products on social media, including promotional images and videos, has been associated with increased e-cigarette use among U.S. adolescents \parencite{king2016exposure, maloney2016does, pokhrel2018social, kim2019effects, wang2019tobacco}.  Instagram, one of the most popular social media platforms among adolescents, is considered the largest source of e-cigarette social media advertisements (cite). E-cigarette stores, distributors and social media influencers - users with large followings who post vaping content on behalf of e-cigarette brands - promote ENDS (Electronic Nicotine Delivery Systems) products on Instagram and other social media \parencite{vassey2020vape}. \\

In August 2018, the U.S. Food \& Drug Administration (FDA) introduced a requirement that e-cigarette advertisements, including social media imagery, contain a prominent warning label that reminds consumers that nicotine is addictive.
The FDA requires that the warning statement appears on the upper portion of the advertisement, occupies at least 20 percent of the advertisement area, and be printed in in conspicuous and legible at least 12-point sans serif (e.g. Helvetica or Arial bold) font size \parencite{fda202020advertising}. Several studies \parencite{vassey2020vape, laestadius2020compliance} evaluated compliance with the 2018 FDA requirements for warning labels on Instagram.
\textcite{vassey2020vape} manually reviewed 2,000 images posted in 2019 and found that only 7\% included FDA-mandated warning statements.  Posts uploaded from locations within the U.S. had the highest prevalence of warning labels, while posts uploaded from other countries were less likely to include warnings. Most of the international posts featured vaping products distributed in the U.S. and would therefore still be subject to compliance with the FDA warning-label regulations \parencite{vassey2020vape}. \textcite{laestadius2020compliance} manually reviewed 1,000 posts collected in late 2018-early 2020 and found that only 13\% included warning statements.  \\

Both studies \parencite{vassey2020vape, laestadius2020compliance} conducted qualitative analysis to assess the presence of warning statements on a small sample size of Instagram posts and used binary classification (presence or absence of a warning statement) without reporting warning labels that were too small or in the wrong place, which would constitute a partial compliance with the FDA requirements. This study addresses the limitation of prior research by developing an automated deep learning image classification capable of quickly and accurately measuring  compliance with the FDA requirements for warning labels on a large sample size. i.e. thousands of images. %
We tested whether advanced deep learning techniques could provide an effective, automated method to track vaping-related posts on Instagram and evaluate compliance with FDA warning label requirements. %

\section{Related work}

Automated analysis of social media content has exploded in popularity in recent years, with relevant comparable studies falling into two major groups: social influence \& marketing focused and health focused. Starting with social influence, which is broader and typically disregards domain, multimodal learning has been deployed in the recent few years for predicting image popularity. Before the rise of deep learning, support vector machines were primarily used, leveraging  basic image features such as color and saturation~\parencite{khosla2014makes}. In recent years, the focus has turned to the use of deep learning to synthesize information and learn feature representations directly from the data. Hu et al. utilize both tags and images~\parencite{hu2016multimodal}, leveraging the Yahoo Flickr Creative Commons 100M dataset. Zohourian and colleagues predicted the number of likes via context information~\parencite{zohourian2018popularity}. De et al. use standard deep neural networks (DNN) to predict future popularity of the Instagram of a lifestyle magazine~\parencite{de2017predicting}. Each of these cases demonstrate the inclusion of different features in regards to metadata, context, time, and social indicators.

However, by far the largest specific domain of interest is related to health. This can be further separated into a) mental health, and b) public health. For instance, multimodal learning has been used to detect depression on Instagram, using a convolutional neural network with synthesizes the aforementioned temporal features~\parencite{chiu2021multimodal}. A related, more general study is integrating text and images to determine the intent of a post~\parencite{kruk2019integrating}, including the promotion of vaping.

Apart from mental health, researchers have also targeted illicit activities related to public health. Yang and Luo used inductive transfer learning, with multi-staged regression using post and account related data~\parencite{yang2017tracking}. The model was a standard DNN but with lower levels shared among different tasks, typical of shared learning tasks, and found it more effective than prior approaches with less modalities. The potential for abuse of social media~\parencite{allem2016importance,allem2017cigarette,allem2018could}, especially in the context of the COVID-19 pandemic~\parencite{chen2020tracking,young2021disrupting}, has also driven the interest in multimodal machine learning. Mackey and colleagues studied suspicious COVID-19 related health products by combining natural language processing and deep learning to produce a binary classifier~\parencite{mackey2020big}. In addition to a useful temporal account of how products related to waves of infection, they demonstrated the efficacy of deep learning in identifying buying/selling intent.

\section{Data and methods}

\subsection{Training data}

The dataset consisted of 4,363 images collected from 3,484 distinct posts.  Our prediction targets consisted of three binary labels: the presence of a correctly placed and sized fully-compliant warning label, a non-fully-compliant warning label, and whether the post promoted vaping or did not. 

\subsection{Evaluation}

The dataset was randomly divided into a 60\% training set, a 20\% validation set for early stopping, and a 20\% test set for evaluation of model performance. This splitting procedure was conducted at the post level rather than the image level, so that multiple images from the same post would all be assigned to the same split. Models were optimized to minimize cross-entropy loss on the binary labels, equivalent to negative log-likelihood loss in logistic regression, and were assessed for accuracy and area under the curve (AUC).

\subsection{Deep learning models}

Our baseline model was a transfer learning-based convolutional neural network (CNN) built in Tensorflow Keras \parencite{chollet2018deep, abadi2016tensorflow}. The output of the image backbone model was passed through a global averaging layer, a dropout of 40\% or 50\% was applied to mitigate overfitting, and then passed to the output layer.  The model was trained with the Adam optimizer \parencite{kingma2014adam} using cross-entropy loss and a batch size of 32 where possible, or 16 when limited by GPU memory (EfficientNet B3).


We tested an EfficientNet model \parencite{tan2019efficientnet} as the primary image processing backbone with pretrained weights from ImageNet \parencite{deng2009imagenet}. To provide benchmark results, we compared EfficientNet to popular alternative architectures that have shown good performance in related research: VGG16 \parencite{simonyan2014vgg}, ResNet50 \parencite{he2016resnet}, Inceptionv3 \parencite{szegedy2016inceptionv3}, and MobileNet \parencite{howard2017mobilenets}. 

In the bias initialization variation, the bias parameter for each output node was initialized to $\log(\textrm{positives} / \textrm{negatives})$ to hasten training convergence with class imbalance \parencite{karpathy2019recipe}.

In the progressive unfreezing variation, the image backbone model was initially frozen and the output head was trained at a relatively higher learning rate to improve the randomly initialized weights, up to 30 epochs with a patience of 2 epochs. Then the last 20\% of layers of the image backbone were unfrozen and the model was trained for up to 30 additional epochs with a reduced learning rate and early stopping with a patience of 3 epochs. Finally the entire image backbone was unfrozen and the model was trained at an even lower learning rate for up to 30 epochs, with a patience of 3 epochs.

\begin{figure}[htbp]
\centering
\includegraphics[width=120mm]{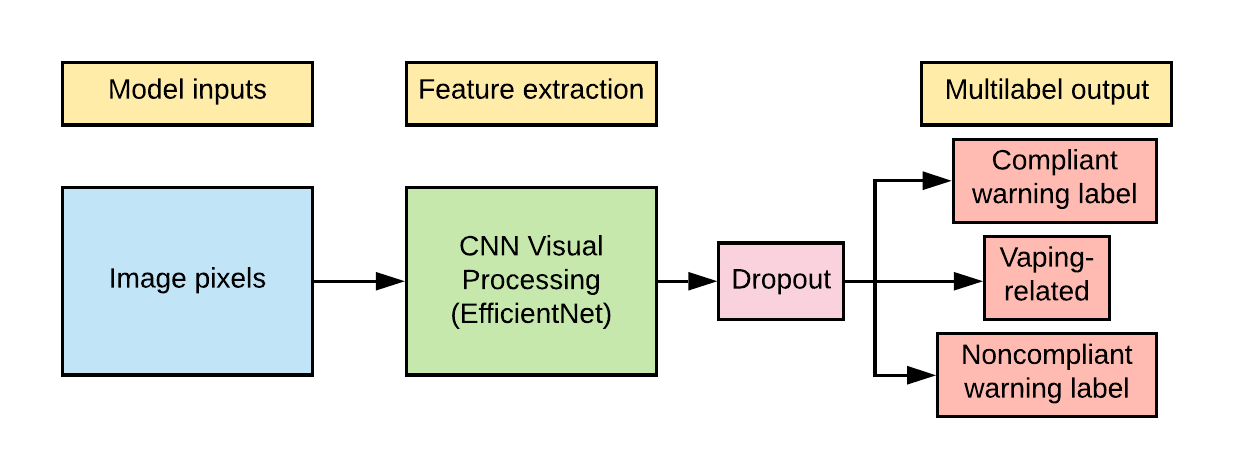}
\caption{\textbf{Image-only multitask model architecture.} }
\label{architecture-multitask}
\end{figure}

The multitask design (Figure \ref{architecture-multitask}) was implemented by including three nodes with sigmoid activation in the output layer, one for each of our binary outcomes. 

Given that the warning label in an Instagram image consists entirely of text, we hypothesized that sufficiently accurate scene text recognition (STR) could increase the accuracy of our model. Scene text recognition is a variant of optical character recognition (OCR) designed to detect and extract arbitrary text in photographs with complex backgrounds, whereas OCR is designed to extract more standardized text from scanned documents or books \parencite{long2021scene}. We used the CRAFT method \parencite{baek2019character} to detect the location of text in the images, followed by a convolutional recurrent neural network (CRNN) model to recognize the specific text characters, which was pretrained on the MJSynth dataset \parencite{Jaderberg14c}.

\section{Results}

Our final model achieved an area under the curve (AUC) and [accuracy] of 0.97 [92\%] on vaping classification, 0.99 [99\%] on FDA-compliant warning labels, and 0.94 [97\%] on non-compliant warning labels, indicating that our sample size and modeling approach was sufficient to achieve excellent predictive accuracy.

We found that the image backbone architecture was the single most important factor affecting predictive performance. Looking at single-task results, we found that a VGG16  visual processing backbone led to an overall cross-entropy loss of 0.53. ResNet50 was comparable at 0.51 loss. By contrast, the simplest version of EfficientNet (B0) achieved a loss of 0.36. This ranking confirmed our expectations, 
mirroring the performance of these models in recent years' benchmarks.

We did not find a benefit from using higher capacity, more complex versions of the EfficientNet architecture. Upgrading from the B0 model (224x244 resolution, 5.3M parameters) to the B3 model (300x300 resolution, 12M parameters) led to an effectively equivalent cross-entropy loss of 0.38. The B5 and B7 versions showed similar results. This may be a reflection of problem complexity, sample size limitations, or perhaps suboptimal training strategy for the larger models. 
Given its computational efficiency, it was advantageous for our speed of experimentation that the lower-parameter B0 version yielded excellent performance.

Initializing the bias weights for each output node to account for class imbalance was beneficial for both lower cross-entropy loss and faster convergence, although a less important consideration than the base architecture with a typical improvement of only 0.02 in cross-entropy.

Many variations that we tested did not show any noticeable performance benefit. A multitask architecture, learning rate finding, freezing of batchnorm layers, and progressize layer unfreezing all yielded no improvements, despite being considered best practices in the deep learning literature. 
These findings reinforce the need to verify that common training practices do show benefit for the particular dataset and task at hand.

\section{Discussion}

Of the three tasks, the identification of FDA-compliant warning labels was easiest for the deep learning models in terms of area under the receiver operating characteristic curve (AUC), even when using only raw pixel information (i.e., no scene text extraction). That is reasonable because the guideline-compliant warning label is featured prominently in the image, making it straightforward for convolutional filters to detect. Determining if an image was promoting vaping proved to be the most difficult task; increasing our sample size and variety of vaping and non-vaping images should be a straightforward way to improve model performance on that task. 

The critical role of the image backbone in driving performance highlights the importance of using a recent backbone architecture when accuracy matters. The rapid innovation in computer vision leads to improved architectures on a yearly basis, or sometimes even more quickly. It also suggests that accuracy results should be interpreted as conditional on the selected backbone architecture, rather than representing a static, final result for a particular dataset. New innovations in computer vision architectures can be reasonably expected to reduce overfitting and increase predictive accuracy over time, even with the same training dataset.

We did not find any images with pseudo-warning text, which would have the initial appearance of being a compliant warning label but that actually contained non-warning language. Such adversarial strategies could lead to false positives in models that rely purely on general convolutional pixels. Incorporation of scene text recognition that is localized to the warning label box would provide robustness to such adversarial strategies, and can be explored in future work. Image augmentation that includes the automatic generation of adversarial labels could also be used to ensure that models are not vulnerable to fake warning labels.

A key limitation of the current study is that we used a single test set for performance evaluation. In future work we plan to use 5-fold cross-validation, facilitating the construction of confidence intervals and eliminating the reliance on a single test set.

\begin{figure}[htbp]
\centering
     \begin{subfigure}[b]{0.45\textwidth}
         \centering
         \includegraphics[width=0.72\textwidth]{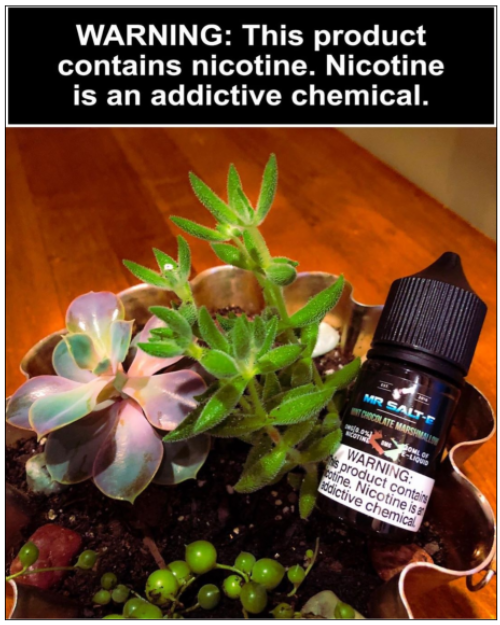}
         \caption{Original image}
     \end{subfigure}
     \hfill
          \begin{subfigure}[b]{0.45\textwidth}
         \centering
         \includegraphics[width=\textwidth]{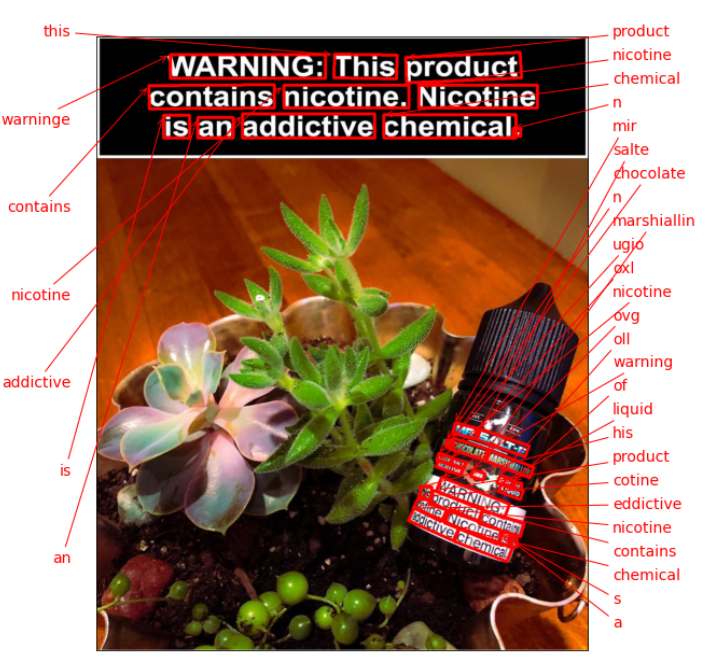}
         \caption{Scene text recognition visualized}
     \end{subfigure}

\caption{\textbf{Example of scene text recognition in an Instagram image.} }
\label{str-example}
\end{figure}

In future work we plan to incorporate the extracted text from scene text recognition into the deep learning architecture, which we have successfully implemented for this dataset. Given our excellent current results, there may not be much remaining benefit for these three tasks by adding the recognized scene text. Nevertheless, such extracted text could prove useful for extracting marketing themes, identifying promoted e-cigarette flavors or brands, and other more granular analysis. An example Instagram image with scene text recognition is shown in Figure \ref{str-example}.

\begin{figure}[htbp]
\centering
\includegraphics[width=\columnwidth]{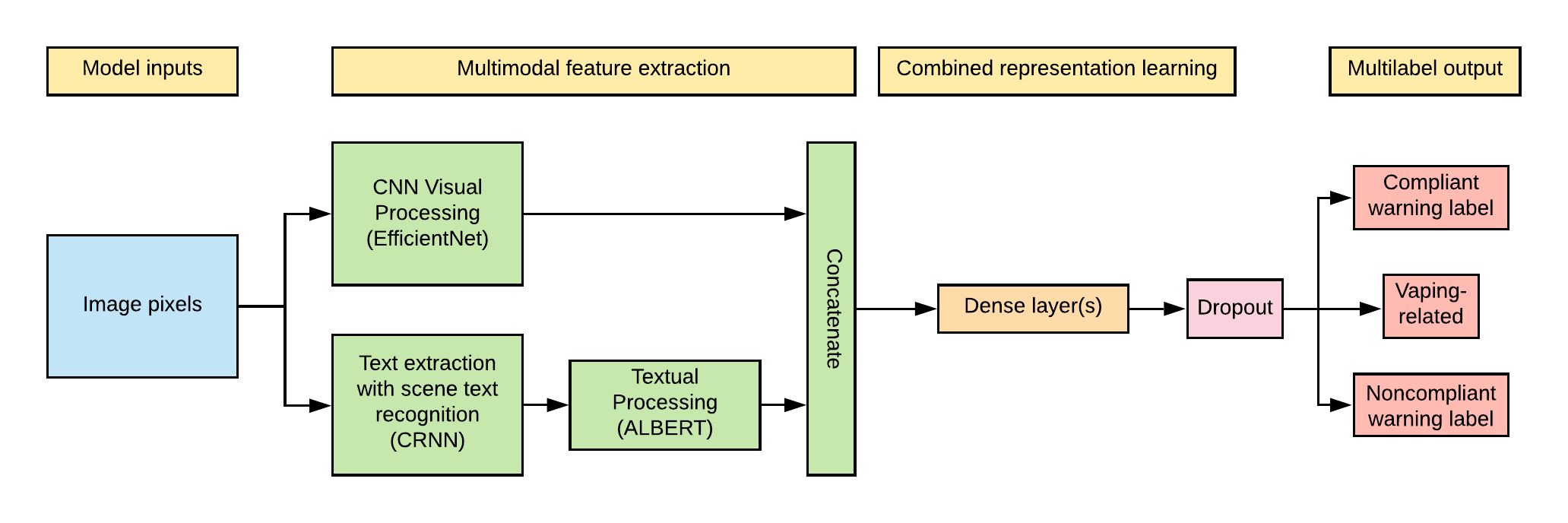}
\caption{\textbf{Image and scene text multitask model architecture.} }
\label{architecture-scene-text}
\end{figure}

The network architecture for our expanded model with scene text recognition is shown in Figure \ref{architecture-scene-text}.

\newpage
\section{Conclusion}

In this study, we demonstrated that deep learning models can successfully identify FDA-compliant e-cigarette warning labels (or lack thereof) on Instagram with high accuracy. The models were further able to determine if a post promoted vaping or did not, and if it contained a non-compliant warning label. In combination, the models allow the automatic tracking of vaping-related posts on Instagram and evaluation of their compliance with FDA marketing policies. Through experimentation across several dimensions, we identified combinations of model architectures and training strategies that were most successful at achieving high accuracy. Future research using deep learning based detection is warranted to evaluate compliance with the FDA warning statements on images and videos on Instagram and TikTok, the most popular social media platforms among teenagers. In particular, the incorporation of scene text recognition has the potential to yield additional granularity for establishing marketing themes and identifying specific brands and flavors.

\section*{Funding}

The authors did not receive any funding for this research.

\newrefcontext[sorting=nyt]
\printbibliography[heading=bibintoc,title={References}]

\newpage

\renewcommand{\thesection}{\Alph{section}}
\setcounter{section}{0}

\end{document}